\documentclass{aa}  

\usepackage{graphicx}
\usepackage{mathtools}
\usepackage{txfonts}
\usepackage{lipsum}
\usepackage{ulem}
\usepackage{caption}
\usepackage{subcaption}         % necessary for continued figures, example in section 3
\usepackage{threeparttable}
\usepackage{mathrsfs}
\usepackage{amsmath}
\usepackage{placeins}
\usepackage{comment}
\usepackage{afterpage}
\usepackage{lscape}             % to rotate a single page table, example in appendix.
\usepackage{placeins}           % useful with \FloatBarrier, to keep 
\usepackage{hyperref}
\hypersetup{
    colorlinks=true,
    linkcolor=blue,
    citecolor=blue,
    urlcolor=green,
    }
\usepackage{natbib}
\begin{document}

%%%%%%%%%%%%%%%%%%%%%%%%%%%%%%%%%%%%%%%%
% if you use custom commands in your title,
% ensure to check your title when submitting!
%%%%%%%%%%%%%%%%%%%%%%%%%%%%%%%%%%%%%%%%
   %\title{A potential faint persistent radio source associated with a nearby fast radio burst}

\title{Fast radio burst -- persistent radio source systems}
\subtitle{II. A faint PRS associated with the nearby FRB 20181030A?}

   \author{D. Pelliciari
          \inst{1},
          G. Bernardi
          \inst{1,2,3},
          B.~Margalit
          \inst{4},
          B.~D.~Metzger
          \inst{5,6},
          C. Nanci
          \inst{7},
          L. Bruno
          \inst{1},
          M. Pilia
          \inst{8},
          L. Beduzzi
          \inst{1,9},
          C. Spingola
          \inst{1},
          C. Stanghellini
          \inst{1},
          P. Esposito
          \inst{10,11},
          A. Geminardi
          \inst{8,10,12},
          M. Giroletti
          \inst{1}
          }

   \institute{
       Istituto Nazionale di Astrofisica (INAF) - Istituto di Radioastronomia (IRA), via Gobetti 101, 40129 Bologna, Italy  
      \and
       Department of Physics and Electronics, Rhodes University, PO Box 94, Makhanda, 6140, South Africa
      \and 
      South African Radio Astronomy Observatory, Black River Park, 2 Fir Street, Observatory, Cape Town, 7925, South Africa 
      \and
       School of Physics and Astronomy, University of Minnesota, Minneapolis, MN 55455, USA
      \and
       Department of Physics and Columbia Astrophysics Laboratory, Columbia University, New York, NY 10027, USA
      \and
       Center for Computational Astrophysics, Flatiron Institute, 162 5th Ave, New York, NY 10010, USA
      \and
     Istituto Nazionale di Astrofisica (INAF) - Osservatorio di Astrofisica e Scienza dello Spazio (OAS) di Bologna, Via P. Gobetti 93/3, 40129, Bologna, Italy
          \and
    Istituto Nazionale di Astrofisica (INAF) - Osservatorio Astronomico di Cagliari (OAC), via della Scienza 5, I-09047, Selargius (CA), Italy
    \and
    Dipartimento di Fisica e Astronomia (DIFA), Universit\`a di Bologna, via Gobetti 93/2, 40129 Bologna, Italy
    \and
    Scuola Universitaria Superiore IUSS Pavia, Palazzo del Broletto, piazza della Vittoria 15, I-27100 Pavia, Italy
    \and
    INAF-Istituto di Astrofisica Spaziale e Fisica Cosmica di Milano, Via Corti 12, I-20133 Milano, Italy
    \and
    Dipartimento di Fisica, Universit\ a di Trento, Via Sommarive 14, I-38123 Povo, (TN), Italy
    \\
    \email{davide.pelliciari@inaf.it}
 }
%   \date{Received September 15, 1996; accepted March 16, 1997}

% \abstract{}{}{}{}{} 
% 5 {} token are mandatory
 
  \abstract
   {Persistent radio sources (PRSs) are the continuum counterparts of fast radio bursts (FRBs), the latter being extragalactic transients of millisecond duration and Jy-level flux density. Up to now only $4$ PRSs securely associated with repeating FRBs are known. An FRB-PRS system is thought to be a flaring magnetar surrounded by a highly magnetized, baryon-loaded nebula. This is consistent with the high RM values ($\approx 10^3-10^5$ rad m$^{-2}$) measured for FRBs having an associated PRS, with the compactness (sub-pc scales) of the latter, and with their high spectral luminosities ($\sim 10^{29}$ erg s$^{-1}$ Hz$^{-1}$ at $1.4$ GHz).}
   {We aim to constrain the size of 20181030A-S1, a new PRS candidate, potentially associated with the repeating FRB 20181030A. The latter is localized with $\sim 1'$ uncertainties in the outskirts of NGC 3252, which is a spiral galaxy at a luminosity distance of $20$ Mpc.}
   {We report very long baseline interferometric (VLBI) observations using the European VLBI Network at $1.7$ GHz of this PRS candidate at an angular resolution of $20$ milliarcseconds.}
   {Our observations reveal the presence of an unresolved radio source (20181030A-S1) at the position of the PRS candidate, confirming its compactness at milli-arcsecond angular scales. A fit to the position of the point-source yields a peak flux density of $280 \pm 30$ $\mu$Jy and a transverse physical size constrained to be $R < 0.5$ pc at $68\%$ confidence level (CL). This flux density converts to a spectral luminosity of $(9 \pm 1) \times 10^{25}$ erg s$^{-1}$ Hz$^{-1}$, $\sim 3$ orders of magnitude lower than confirmed PRSs, making 20181030A-S1 the closest and faintest PRS candidate known. Its low luminosity and modest rotation measure are consistent with the $L_\nu-$rotation measure (RM) relation followed by confirmed FRB--PRS systems, supporting a common physical origin in magnetar-powered nebulae.}
   {We show how a magnetized wind nebula powered by an initially weak ($B_\star \simeq 10^{15}$ G) and young ($t_{\rm age} \simeq 15 - 150$ yrs) magnetar can account for both the observed spectral luminosity and RM of the system. Other possible origin scenarios for 20181030A-S1, in the case in which it is unrelated to the FRB source, are also discussed.}


   
\titlerunning{A new potential persistent radio source compact at pc scales}
\authorrunning{Pelliciari et al.}
   \maketitle
    \nolinenumbers
%
%-------------------------------------------------------------------

\section{Introduction}
Fast Radio Bursts (FRBs) are intense, millisecond-duration radio pulses that originate predominantly from extragalactic distances \citep[see, e.g.,][for a review on the topic]{Zhang22_rev}. Approximately $3600$ different FRBs have been cataloged to date \citep{CHIMECat2}, with a subset $-$ approximately $2\%$ $-$exhibiting repetitive behavior. Nowadays, it is still unclear whether repeaters represent a distinct physical class among the whole FRB population \citep[see, e.g.,][]{BeniaminiKUmar25} and statistically it is plausible that some one-offs will eventually repeat with low bursting activity \citep[see, e.g.,][]{James23}. The most commonly invoked progenitors for FRBs are magnetars, i.e. neutron stars powered by the decay of strong ($10^{14-15}$ G) magnetic fields  \citep[e.g.,][]{DuncanThompson, ThompsonDuncan, Popov13, Lyubarski14}. This scenario, motivated by theoretical works \citep[e.g.][]{Popov10,Popov13,Lyubarski14,Metzger17,Beloborodov19,Liubarsky20,Sobacchi22}, is supported by indirect observational evidences \citep[see, e.g.,][]{Michilli18,Kramer24,Nimmo25,Shah2026} and was further bolstered by the 2020 discovery of an X-ray burst simultaneous with an FRB-like radio signal from the Galactic magnetar SGR J1935+2154 \citep{Israel16,CHIME20b,Bochenek20a,Mereghetti20,Tavani21,Ridnaia20}.

The milli-arcsecond localization of the repeating FRB 20121102A (R1) revealed its association with a compact \citep[$<1$ pc;][]{Marcote17} persistent radio source (PRS) in a low-metallicity, dwarf galaxy at $z=0.193$, exhibiting a high luminosity of $L_\nu = 2.8 \times 10^{29}$ erg s$^{-1}$ Hz$^{-1}$ and a flat spectral index $\alpha = -0.27$ \citep{Chatterjee17,Marcote17}, where $S(\nu) \propto \nu^\alpha$ is the source flux density, both inconsistent with a star forming region \citep[e.g.][]{Klein18,Piro21}. Given also the high ($\sim 10^5$ rad m$^{-2}$) rotation measure (RM) of R1 bursts \citep{Michilli18}, this PRS may be a strongly-magnetized, synchrotron-emitting nebula powered by a young magnetar, where relativistic winds inflate a bubble that interacts with the interstellar medium, producing the observed persistent emission \citep{margalitmetzger18}. Alternative models invoke accretion onto compact binaries or black hole systems, where the PRS arises from jet activity or wind interactions \citep{Sridhar24}.

Subsequent observations identified three additional PRSs associated with repeating FRBs, i.e. FRBs 20190520B \citep{Niu21,Bhandari23b}, 20240114A \citep{Bhusare24,Bruni24} and recently 20190417A \citep{Ibik24,Moroianu25}, all confirmed to be compact on parsec scales and precisely linked with the their associated FRBs, given also the high (sub-arcsecond level) positional accuracy of the latter.

These persistent sources present various similarities with the PRS associated with R1, i.e. a flat spectral index \citep[see][]{Bruno26},  bursts with high and time-variable RMs \citep[e.g.,][]{McKinven23,Moroianu25} and their localization in low-metallicity dwarf galaxies \citep[see][and references therein]{Moroianu25}. These shared traits may suggest the presence of a distinct FRB sub-population arising from young, magnetized progenitors in low-metallicity environments, hosting bright PRSs.

Recently, a PRS candidate has been reported at arcsecond angular scales for the repeating FRB 20181030A via Very Large Array (VLA) observations at $1.5$ GHz \citep{Ibik24}. This FRB, first discovered in 2018 by the Canadian Hydrogen Intensity Mapping Experiment \citep[CHIME;][]{chime18}, has been associated with the nearby star-forming spiral galaxy NGC 3252 \citep{Bhardwaj21}, located at a redshift of $z = 0.00385 \pm 0.00002$ \citep[$\sim20$ Mpc;][]{Masters14, Tully16}. In particular, of all the FRB 20181030A host candidates, NGC 3252 is the only one with a redshift compatible with the dispersion measure (DM) of the FRB source, i.e. ${\rm DM} = 103.5$ pc cm$^{-3}$ \citep{chime18, Bhardwaj21}. The continuum radio source reported in \citet{Ibik24} (20181030A-S1, hereafter, according to the nomenclature adopted by the authors), overlaps with a spiral arm of the FRB host galaxy but neither the compactness of the source at pc scales, nor the secure association with the FRB has been established yet. 

In this work, we present new European VLBI Network (EVN) observations aimed at constraining the compactness of 20181030A-S1 at milli-arcsecond scales, providing further support for its classification as a PRS, similar to the confirmed ones.  In \S\ref{sec: observations_and_reduction} we describe the observations conducted along with the applied data reduction. We present our results in \S\ref{sec: results} and discuss their implications in \S\ref{sec: implications}. Finally, we draw our conclusions in \S\ref{sec: conclusions}.

\section{Observations and data reduction}\label{sec: observations_and_reduction}

\subsection{EVN+eMERLIN observations}
We observed FRB 20181030A on September 17$^{\rm th}$ 2025, from $4^{\rm h}$ till $8^{\rm h}$ UT (project ID ep134d, PI: Pelliciari) using a total of eleven antennas. These are: Effelsberg (Ef), Jodrell Bank (Jb), Westerbork single-dish RT1 (Wb), Medicina (Mc), Onsala (O8), Tianma (T6), Irbene (Ir), Cambridge (Cm), Knockin (Kn), Pickmere (Pi) and Defford (De). However, strong wind speeds at O8 site prevented observations, and no fringes were obtained for enhanced Multi Element Remotely Linked Interferometer Network (eMERLIN) antennas Cm, Kn and Pi. This resulted in seven operating antennas. 

Observations were carried out at a central frequency of $1.65$~GHz, with EVN antennas providing $4 \times 32$ MHz subbands and eMERLIN dishes $2 \times 64$ MHz subbands. 
The phase center was at ${\rm RA}_{\rm J2000} = 10^{\rm h} \, 34^{\rm m} \, 14\fs25$ and ${\rm Dec}_{\rm J2000} = +73^\circ \, 45' \, 4\farcs10$, consistent with the VLA localisation of 20181030A-S1  \citep{Ibik24}. The source 3C 147 was observed as a bandpass/fringe-finder calibrator. The source J1027+7428 was observed as a phase-reference calibrator (angular separation from the target: $0\fdg86$), alternated to the target, for $2.5$~minutes for each time slot. The target source was observed for a total of $2.8$~hours. We observed J1056+7011 as a further calibrator to assess the goodness of the calibration procedure.

Visibilities were generated using the SFXC software correlator \citep{Keimpema15} at JIVE, with a $2$~s integration time. Each subband was channelized into $64$~channels, $500$~kHz each. The Astronomical Image Processing System \citep[\textsc{aips;}][]{Griesen03} and the Common Astronomy Software Applications package \citep[\textsc{casa};][]{McMullin07} were used for the data reduction. We used the \textsc{fitld} task in \textsc{aips} to load the correlated visibilities in the standard FITS-IDI format. Gain curves and system temperature measurements collected at each station during the observations were applied to the visibilities for the amplitude calibration, followed by parallactic angle correction and RFI flagging. We also corrected for ionospheric Faraday rotation and dispersive delay exploiting maps of total electron content (TEC), using the \textsc{tecor} task in \textsc{aips}. Finally, we solved for the complex bandpass and we exported visibilities to UVFITS format using the \textsc{fittp} task in \textsc{aips}.

\begin{figure}
    \centering
    \includegraphics[width=1.0\columnwidth]{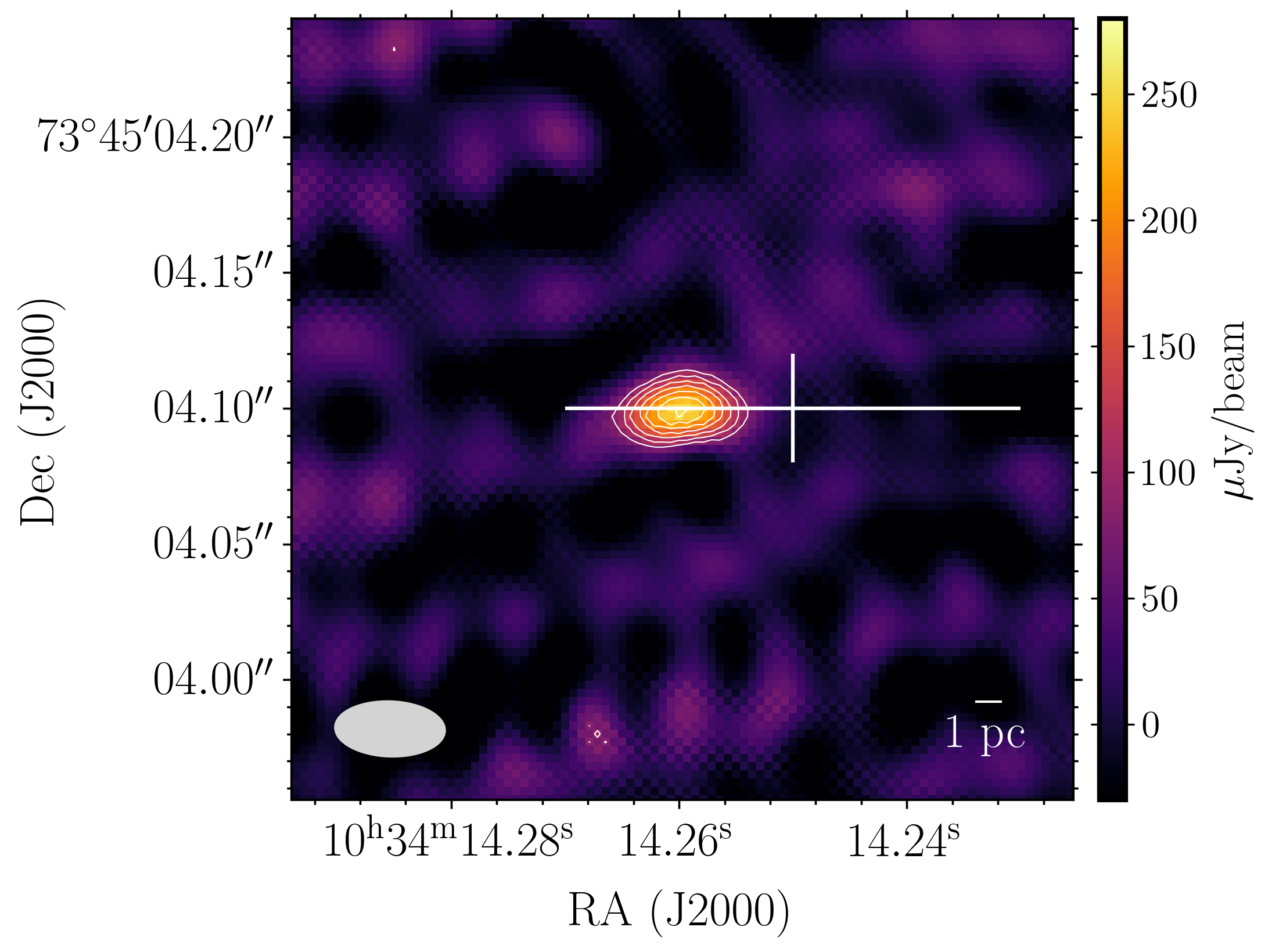}
    \caption{EVN cleaned image of 20181030-S1. The contour levels range from three to nine times the rms noise level $\sigma \simeq 28$ $\mu$Jy beam$^{-1}$. The $0\farcs040 \times 0\farcs020$ synthesized beam of EVN observations is represented as a filled gray ellipse at the bottom left corner of the image. The white cross is centered on the position of 20181030A-S1 as previously reported in \citet{Ibik24} with $1.4$ GHz VLA observations at arcsecond angular resolution, with extensions being the $1\sigma$ uncertainties on its centroid. The small white bar at the bottom right corner shows a $1$ pc extent at $z = 0.00385$ (the redshift of the FRB host galaxy).}
    \label{fig: PRS_img}
\end{figure}

Visibilities were then converted from UVFITS to measurement set (MS) format using the \textsc{casa} \textsc{importuvfits} task and flagged using \textsc{flagdata} task after a visual inspection. Initially, the calibration source 3C 147 was used as a fringe finder to solve for the phases and delays for each spectral window (i.e. single-band delay), due to instrumental effects. These solutions were then applied to the secondary calibrator J1027+7428, which we used to solve for delays and delay rates across all the secondary calibrator scans combining all the spectral windows (i.e. global fringe fitting). Calibration was followed by flagging and the procedure was iterated over until calibration solutions converged. All calibration solutions were then applied to the target and the check source.

After the calibration procedure described above, the visibilities corresponding to the 20181030A-S1 observations were Fourier transformed into a residual (dirty) image with field of view (FoV) $21'' \times 21''$, using the natural weighting scheme as routinely done in previous PRS observations conducted with the EVN \citep{Marcote20, Kirsten22glob, Nimmo22, Bhandari23b, Hewitt24}. We used the multifrequency synthesis algorithm to combine together all the spectral windows. The dirty image clearly shows a point source at the image center and no other sources across the field of view.

We performed a deconvolution down to a $28~\mu$Jy~beam$^{-1}$ root mean square (rms) noise and obtained a clear detection at signal-to-noise ratio (S/N) $\sim 9$ of 20181030A-S1. Fig. \ref{fig: PRS_img} shows the cleaned image of the target field. We fitted a two dimensional Gaussian to the cleaned source and found its position to be:

\begin{equation*}
\begin{aligned}
    &\alpha({\rm J2000}) = 10^{\rm h} 34^{\rm m}  14\fs230 \pm 5.9 \ {\rm mas}, \\
    &\delta({\rm J2000}) = +73^\circ 45' 4\farcs100 \pm 6.0\ {\rm mas},
\end{aligned}
\end{equation*}

referenced to the international celestial reference frame (ICRF). Following \cite{Bhandari23b} and \cite{Moroianu25}, the quoted positional uncertainties are the quadrature sum of multiple contributions:  the fitting error derived
from the synthesized beam shape and ${\rm S/N} \sim 9$ of the detection ($\Delta \alpha = 4.4$ mas, $\Delta \delta = 2.2$ mas); the uncertainties in the absolute positions of the phase calibrator (J1027+7428; $\Delta \alpha = 0.17$ mas, $\Delta \delta = 0.13$ mas) and check source
(J1056+7011; $\pm 0.03$ mas); the check source positional offset ($\Delta \alpha = 4.0$ mas, $\Delta \delta = 5.2$ mas); and an estimate of the frequency-dependent shift in the
phase calibrator and check source positions, conservatively $\pm 1$ mas for each \citep{Moroianu25}. Our observations improve by a factor $\approx 14$ the precision on the source position, previously reported in \cite{Ibik24}.

\subsection{Archival observations}
The field of FRB 20181030A is covered by the Third Data Release of the LoFAR Two Meter Sky Survey \citep[LoTSS DR3;][]{Shimwell26} at $120$-$168$ MHz and by the Very Large Array Sky Survey \citep[VLASS;][]{Lacy20} at 2-4 GHz frequencies. We measure an rms noise of $73$ $\mu$Jy beam$^{-1}$ and $118$ $\mu$Jy beam$^{-1}$ at 144 MHz and 3 GHz, respectively. These converts to $1\sigma$ spectral luminosity thresholds of $2.6 \times 10^{25}$ erg s$^{-1}$ Hz$^{-1}$ at $144$ MHz and $4.1 \times 10^{25}$ erg s$^{-1}$ Hz$^{-1}$ at $3$ GHz, considering the luminosity distance $D_L = 20 \pm 5$ Mpc and the redshift $z = 0.00385 \pm 0.00002$  of NGC 3252 \citep{Tully16, Masters14}. We use these observations to characterize the broadband spectrum of 20181030A-S1 in \S\ref{sec: spec}.

\section{Results}\label{sec: results}
\subsection{20181030A-S1 is compact at sub-pc scales}\label{sec: sec1}

The flux density of the source is found to be $280 \pm 30$~$\mu$Jy. Note that this value can vary by a $15\%$, i.e. the typical absolute flux scale uncertainty in VLBI observations \citep[e.g.][]{Bhandari23b}. The measured flux density is $\approx 25\%$ lower with respect to the one reported for the same source at arcsecond angular scales \citep[i.e. $400 \pm 20$ $\mu$Jy;][]{Ibik24}. We attribute this difference to an extended component not ascribable to the (potential) PRS, that our observations are resolving out. At the luminosity distance of the FRB host galaxy, i.e. $20 \pm 5$ Mpc, the measured flux density converts to a spectral luminosity of $L_\nu = (9.7 \pm 0.1) \times 10^{25}$ erg s$^{-1}$ Hz$^{-1}$. This value of spectral luminosity is obtained by considering a luminosity distance of $D_L \approx 17.3$ Mpc, which is computed from the redshift of NGC 3252, assuming a Planck 2018 cosmology \citep{Planck20}, and it is consistent at 1$\sigma$ confidence level (CL) with the value reported in the literature, i.e. $D_L = 20 \pm 5$ Mpc \citep{Bhardwaj21}. As already anticipated in \citet{Ibik24}, this luminosity is $\sim 3$ orders of magnitude lower with respect to the ones reported for the other confirmed PRSs, which show $L_\nu \approx 10^{29}$ erg s$^{-1}$ Hz$^{-1}$ \citep[see, e.g., Table 2 in ][]{Moroianu25}. 

\begin{figure}
    \centering
    \includegraphics[width=0.9\columnwidth]{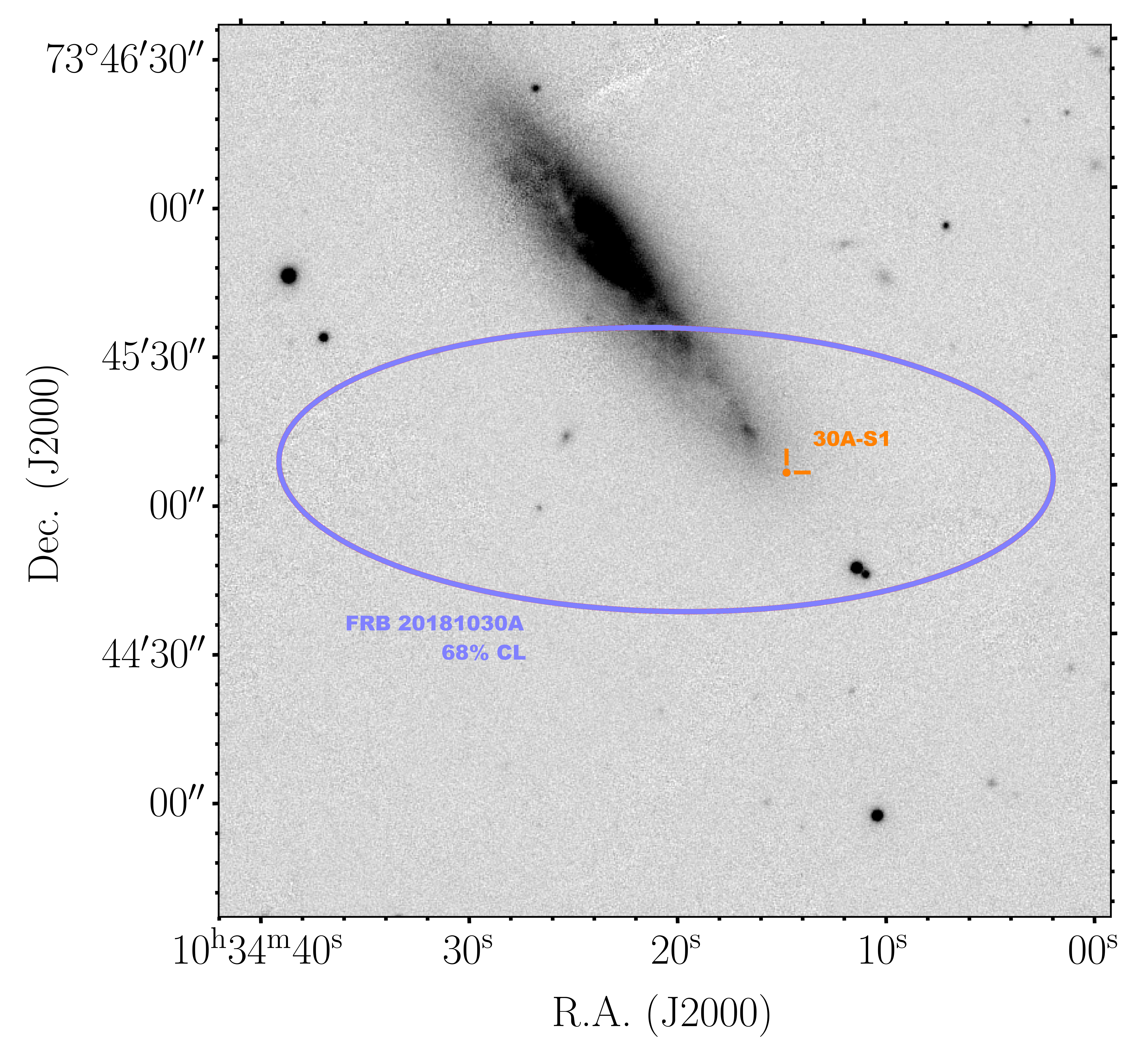}
    \caption{Panoramic Survey Telescope and Rapid Response System \citep[Pan-STARRS;][]{Flewelling20} optical image (z filter) of the field around FRB 20181030A, of which $68\%$ localization uncertainties \citep{Bhardwaj21} are shown as a blue ellipse. The position of 20181030A-S1 is indicated with an orange point. The spiral galaxy in the upper part of the image is NGC 3252, i.e. the most probable host galaxy associated with FRB 20181030A \citep{Bhardwaj21}.}
    \label{fig: loc_PRS}
\end{figure}

Although the coordinates we found for 20181030A-S1 are consistent with the position of the same PRS candidate as reported via VLA observations \citep[see Fig. \ref{fig: PRS_img};][]{Ibik24}, we note that the positional uncertainty ellipse of FRB 20181030A shown in Fig. 6 of \cite{Ibik24} is inconsistent with the localization region reported by \cite{Bhardwaj21}. Henceforth, we show the position of 20181030A-S1 in the field around NGC 3252, along with the corrected $1\sigma$ localization uncertainties of FRB 20181030A \citep[as reported in][]{Bhardwaj21}, in Fig. \ref{fig: loc_PRS}.

Following \cite{Bhandari23b} and \cite{Moroianu25}, we estimate the source angular extension by fitting a Gaussian directly on the observed visibilities, using the \textsc{modelfit} routine of the \textsc{difmap} package \citep{Shepherd1997}. We obtain an upper limit of $12$ mas for the source major axis full width at half maximum (FWHM). The latter translates into an upper limit for the transverse physical size of the source, that must be $R \leq 0.5$ pc\footnote{For this calculation, we considered cosmological parameters as in \citep{Planck20}, although at the low redshift of the source the impact of a different cosmology is negligible.} at 68$\%$ CL. This proves that 20181030A-S1 is compact at sub-arcsec scales. We note that this is the second most stringent constraint on the physical size of a PRS, considering that $R \leq 0.7$ pc for R1 \citep{Marcote17}, respectively. The other confirmed PRSs have shallower constraints, i.e. $R \leq 4$ pc \citep[FRB 20240114A;][]{Bruni24}, $R \leq 9$ pc \citep[R1-twin;][]{Bhandari23b} and $R \leq 23$ pc \citep[FRB 20190417A;][]{Moroianu25}, given also their higher redshift with respect to FRB 20181030A.

\subsection{Extending the relation between FRB RM and PRS spectral luminosity}

The current, most invoked physical interpretation of a PRS is a magnetized synchrotron-emitting nebula surrounding a central magnetar \citep{margalitmetzger18,Rahaman25}, although other interpretations as hyper-nebulae surrounding a binary BH are viable \citep[e.g.][]{Sridhar24}. Regardless of the central engine, \cite{Yang20,Yang22} argued that if the region from which the radio persistent emission originates is the same one that Faraday rotates the bursts from the central FRB source, then a linear relationship between the FRB RM and the PRS spectral luminosity is expected:

\begin{equation}\label{eq: RM_Lnu}
\begin{aligned}
    L_\nu &= \frac{64\pi^3}{27} \zeta_e \gamma_{\rm th}^2 m_e c^2 R^2 |{\rm RM_{\rm src}}|\\
    &\approx 5.7 \times 10^{28}\ {\rm erg\ s^{-1}\ Hz^{-1}}\ \zeta_e \gamma_{\rm th}^2\ \Biggl(\frac{R}{0.01 \ {\rm pc}} \Biggr)^2 \Biggl(\frac{{\rm |RM_{\rm src}|}}{10^4\ {\rm rad\ m^{-2}}} \Biggr) ,
\end{aligned}
\end{equation}

where $R$ is the physical scale of the PRS in units of pc, $\zeta_e$ is the fraction of electrons that radiate synchrotron emission in the GHz band, $\gamma_{\rm th}$ is the Lorentz factor of the thermal electron component \citep[see ][for a detailed description of these parameters]{Yang22}. \cite{Bruni23,Bruni24,Ibik24} showed that the confirmed PRSs follow this relationship, although a large range is considered for the composite normalization parameter $\zeta_e \gamma_{\rm th}^2 (R/0.01 \ {\rm pc})^2 \in (0.1, 10)$. 

The RM of FRB 20181030A has been measured to be ${\rm RM}_{\rm obs} \simeq 37$ rad m$^{-2}$ \citep{McKinven23}, with a Milky Way (MW) contribution of $\approx -19 \pm 6$ rad m$^{-2}$ \citep{Hutschenreuter21}. Henceforth, the RM contribution at the source reference frame is ${\rm RM}_{\rm src} = ({\rm RM}_{\rm obs} - {\rm RM}_{\rm MW}) \times (1+z)^2 \sim ({\rm RM}_{\rm obs} - {\rm RM}_{\rm MW}) \approx 57 \pm 7$ rad m$^{-2}$. We note that this is $\sim 8$ times less than the confirmed PRS having the lowest RM, i.e. the one associated with FRB 20240114A \citep{Tian24_14A, Bruni24}. This makes FRB 20181030A an interesting test case to assess its location on the RM--$L_\nu$ relation and to explore whether it extends it toward lower RM values.

\begin{figure}
    \centering
    \includegraphics[width=1.0\columnwidth]{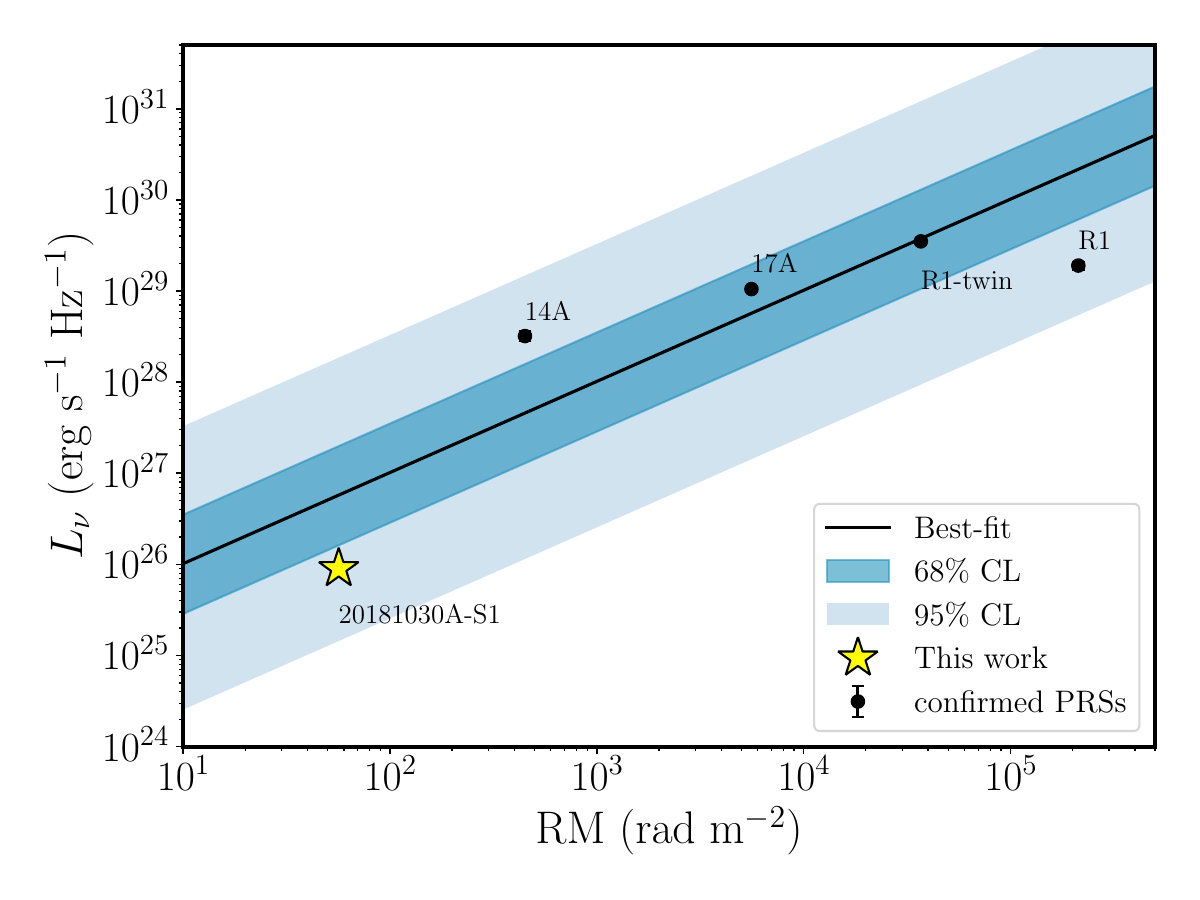}
    \caption{The RM--$L_\nu$ relation for confirmed PRSs (black points). The latter are FRBs 20240114A (14A), 20190417A (17A), 20190520B (R1-twin) and 20121102 (R1). The solid black line represents the best-fit model obtained by fitting Eq. \ref{eq: RM_Lnu} only considering confirmed PRSs, while shaded light and dark blue areas represent the associated 68\% and 95\% CLs. 20181030A-S1 is marked as a yellow star.}
    \label{fig: RML_rel}
\end{figure}

To this end, we considered the sample of confirmed PRSs reported in the literature and performed a Bayesian fit of the RM--$L_\nu$ relation. We parametrized Eq. (\ref{eq: RM_Lnu}) in terms of a single free normalization parameter $A = \zeta_e \gamma_{\rm th}^2 (R/0.01\ {\rm pc})^2$, considering a flat prior distribution in the range $A \in (10^{-5}, 10^3)$. Given the large intrinsic scatter observed in the PRS population, we adopt a log-normal likelihood to model the distribution of $L_\nu$. This choice prevents an over-weighting of PRSs with high RM, which would otherwise dominate the fit because RM spans nearly three orders of magnitude, while the uncertainties on $L_\nu$ (propagated from the measured PRS radio flux densities) are comparatively small. Moreover, to account properly for the large dispersion among the data points (see Fig. \ref{fig: RML_rel}), we include an intrinsic scatter term, $\sigma_{\rm int}$, which enters only through the likelihood function (see Appendix \ref{app:A} for details of the fitting procedure). After verifying the convergence of the MCMC chains, we marginalise over $\sigma_{\rm int}$, treating it as a nuisance parameter. From the MCMC fit, we obtained $A = 1.8^{+3.8}_{-1.3}$ (with uncertainties at 1$\sigma$ level) as the best-fit value for the normalization of the RM--$L_\nu$ relation. 

Fig. \ref{fig: RML_rel} shows the resulting best-fit to the RM--$L_\nu$ relation obtained by considering only the confirmed PRSs, together with the associated 68\% and 95\% CL intervals. Notably, the measured values of ${\rm RM}_{\rm src}$ and $L_\nu$ for 20181030A-S1 are consistent with the best-fit model, although only at 95\% CL. This indicates that, despite its significantly lower RM compared to the confirmed PRS population, 20181030A-S1, if confirmed to be securely associated with FRB 20181030A, follows the same RM--$L_\nu$ relationship, extending the latter to low-RM values and low spectral luminosities. As already discussed, the best-fit obtained in this work considers only confirmed PRSs, and it is based on a completely linear $L_\nu \propto {\rm RM}$ relation. We investigate the possibility of a non-linear RM-$L_\nu$ relation in a companion work (Pelliciari et al. in prep.), in which we perform a bayesian MCMC fit of a large sample of 49 FRB sources with known RM and a measure (or an upper limit) for their persistent spectral luminosity.

\subsection{20181030A-S1 radio spectrum}\label{sec: spec}

We find low-significance radio emission (S/N $\sim 4$) in both LoTSS DR3 and VLASS (see Fig. \ref{fig:LoTSS_VLASS}). However, given the much lower angular resolution of these surveys ($\sim6''\times6''$ for LoTSS and $\sim3.3''\times2.1''$ for VLASS) compared to our EVN observations, any detections should be interpreted with caution, as they may include both compact and diffuse emission. In particular, the LoTSS field is populated by several faint features aligned with the host galaxy disk, making it difficult to disentangle any compact counterpart from diffuse galactic emission. At these angular scales and signal-to-noise ratios, the measured LoTSS flux density should therefore be regarded as a conservative upper limit to the compact emission associated with 20181030A-S1.

In LoTSS DR3, a faint radio feature is detected at a position consistent with the coordinates of 20181030A-S1. We conservatively adopt a flux density of $300 \pm 70,\mu$Jy at a central frequency of 144 MHz, while noting that part of this emission may arise from diffuse radio emission associated with the host galaxy. In VLASS, the source appears unresolved and has a flux density of $500 \pm 120\ \mu$Jy at 3 GHz. These convert to the following spectral luminosities $L_{\rm 144\ MHz} = (1.0 \pm 0.3) \times 10^{26}$ erg s$^{-1}$ Hz$^{-1}$, $L_{\rm 1.4\ GHz} = (1.40 \pm 0.07) \times 10^{26}$ erg s$^{-1}$ Hz$^{-1}$ and $L_{\rm 3\ GHz} = (1.8 \pm 0.4) \times 10^{26}$ erg s$^{-1}$ Hz$^{-1}$, for LoTSS, VLA and VLASS observations, respectively. Recently, \cite{Bruni26} reported the non-detection of 20181030A-S1 at milli-arcsecond scales via EVN observations at 5 and 8 GHz. The corresponding upper limits on the spectral luminosity are $L_{5\,\rm GHz} \leq 3.8 \times 10^{25}$ erg s$^{-1}$ Hz$^{-1}$ and $L_{8\,\rm GHz} \leq 7.2 \times 10^{25}$ erg s$^{-1}$ Hz$^{-1}$.

The radio spectrum of 20181030A-S1 is shown in Fig. \ref{fig:model_SEDs}. We use the available flux density measurements to investigate the scenario of a magnetar wind nebula (MWN), following the framework presented in \cite{margalitmetzger18}. Given the presence of deep upper limits at high frequencies, a radio spectrum described by a single-power law in the $144$ MHz -- $8$ GHz frequency range is ruled out. Considering only VLBI observations, a steep spectrum is implied in the $1.7-5$ GHz range, i.e.  $\alpha \leq -1.1$ at 95$\%$ CL, consistent with the reported upper limit in \cite{Bruni26}.

Overall, the radio spectrum of 20181030A-S1 eappers to peak at frequencies of a few hundred MHz to $\sim1$ GHz and declines steeply at higher frequencies. This conclusion is primarily driven by the robust GHz measurements and high-frequency VLBI upper limits, while the low-frequency LoTSS point should be regarded as tentative because of the possible contribution from diffuse host-galaxy emission.

\section{Implications on the origin of 20181030A-S1}\label{sec: implications}

In the following, we discuss the implications that the results presented in \S\ref{sec: results} have on the origin of 20181030A-S1. We first explore the scenario in which 20181030A-S1 is a genuine PRS associated with FRB 20181030A, and examine the consequences for the population of FRB-PRS systems and their underlying physical mechanisms. In particular we will consider a MWN for the PRS model \citep{margalitmetzger18}. We then discuss alternative interpretations of 20181030A-S1 in the case it is unrelated to the FRB in \S\ref{sec: altern}.

\subsection{Magnetar wind nebula}\label{sec: modeling}

\begin{figure}
    \centering
    \includegraphics[width=1.0\columnwidth]{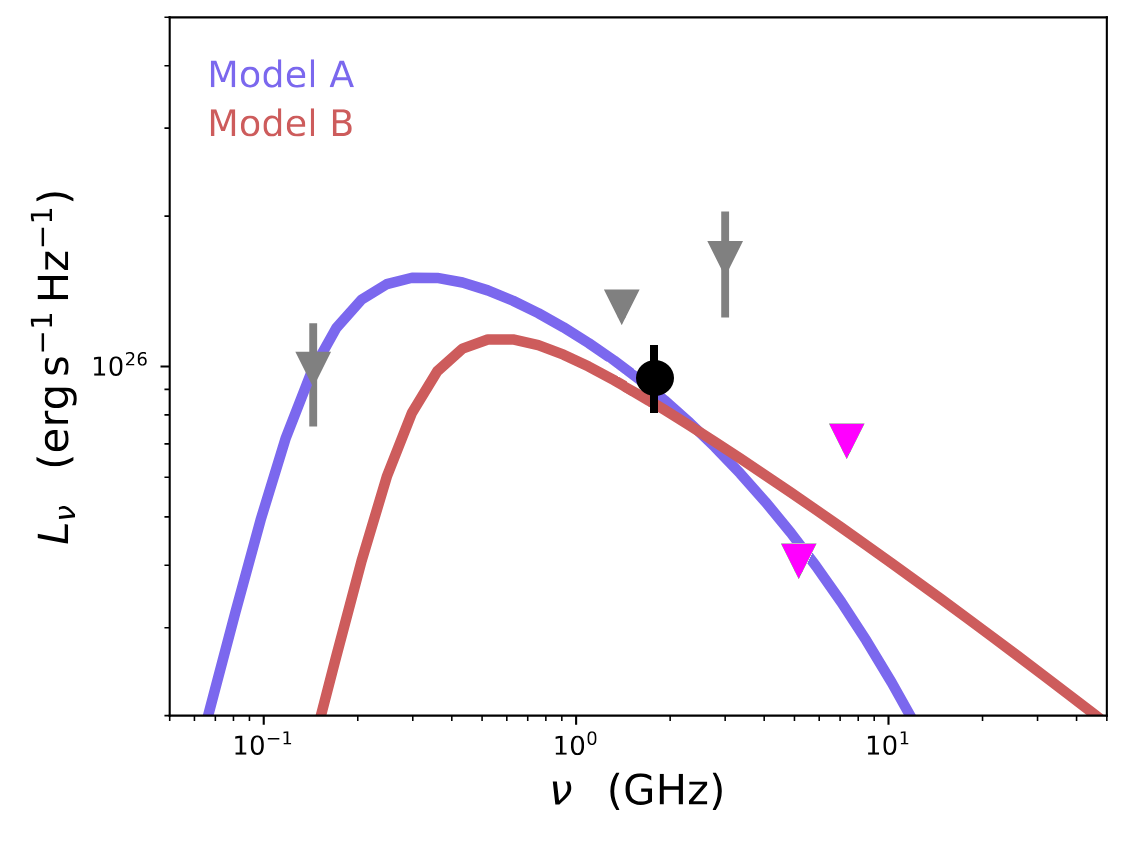}
    \caption{Radio spectrum of two nebula models at time $t=t_{\rm age}$ when the predicted model RM equals the measured ${\rm RM} = 57\,{\rm rad \, m}^{-2}$ of FRB 20181030A. 
    The two models simultaneously match the 1.7 GHz EVN detection of 20181030A-S1 (black point) and the RM. The models are constructed following the \cite{margalitmetzger18} formalism and are described in detail in \S\ref{sec: modeling}.
    For the purposes of this figure, we treat the PRS detections at other frequencies as upper limits due to possible contamination from extended emission beyond the PRS (gray triangles; see \S\ref{sec: spec}). Purple downward triangles represent the non-detection of the 20181030A-S1 at 5 and 8 GHz \citep{Bruni26} with the EVN.}
    \label{fig:model_SEDs}
\end{figure}

We simultaneously model the RM and PRS emission produced by 20181030A-S1 within the \cite{margalitmetzger18} magnetar-nebula model.
Due to the numerical nature of the modeling and computational limitations, we do not attempt formal model fitting or parameter estimation. However we illustrate that the nebula-model is capable of matching the data by showing two different hand-chosen models that do so.
Following the notation of \cite{margalitmetzger18}, Model A was chosen to have:
$E_{B_\star} = 6 \times 10^{47}\,{\rm erg}$, $\alpha = 1.3$, $v_{\rm n} = 1.1 \times 10^8 \,{\rm cm \, s}^{-1}$, and $\chi = 0.4\,{\rm GeV}$.
Here $E_{B_\star}$ is the total magnetic free energy of the magnetar, $\alpha$ controls the magnetar's energy-injection decay rate, $v_{\rm n}$ is the average nebula expansion velocity (here assumed to be constant; see \citealt{Margalit19} for an extension of this treatment), and $\chi$ is the mean energy per particle at injection.
Model B instead has:
$E_{B_\star} = 2 \times 10^{49}\,{\rm erg}$, $\alpha = 1.1$, $v_{\rm n} = 6 \times 10^6 \,{\rm cm \, s}^{-1}$, and $\chi = 10\,{\rm GeV}$.
For both models, the magnetization $\sigma$ and initial time $t_0$ were fixed at $\sigma = 0.1$ and $t_0 = 0.2\,{\rm yr}$.

The particle distribution, density, and magnetic field were evolved according to the framework described in \cite{margalitmetzger18}.\footnote{
Model B was evolved using a semi-analytic calculation which accurately approximates results of the full numerical calculation.
} 
Using these quantities, the RM and synchrotron luminosity (at different frequencies) can be calculated at different times in the nebula's evolution. 
The time at which the model RM equals the observed RM for 20181030A-S1 defines the current `age' of the nebula, i.e. ${\rm RM}(t_{\rm age}) = 57\,{\rm rad \, m}^{-2}$. This implies ages of 14.8 yr and 184.5 yr for Models A and B, respectively. These serve as examples of both young and old nebulae.

Figure~\ref{fig:model_SEDs} shows radio spectra predicted by Models A and B at times $t_{\rm age} = 14.8\,{\rm yr}, \,185\,{\rm yr}$ (respectively) at which the model RM matches the observed RM of the source.
In both cases the model PRS intersects the 1.7 GHz EVN observation, indicating that the models are capable of reproducing both the observed RM and PRS luminosity at a given epoch. Treating the LoTSS, VLA, and VLASS detections as upper limits due to possible contamination by extended emission (see \S\ref{sec: spec}), the models remain broadly consistent with the available constraints, including the recent EVN non-detections at 5 and 8 GHz, although the latter favour models with a more rapidly declining high-frequency spectrum.
Finally, we note that both models additionally satisfy the EVN size constraint derived in \S\ref{sec: sec1}. Model A predicts a nebula size of $R = 0.011\,{\rm pc}$ at the current epoch ($t=t_{\rm age}$), while Model B predicts $R = 0.017\,{\rm pc}$.

In comparison to analogous models for FRB 20121102A \citep{margalitmetzger18}, the two models for 20181030A-S1 described above imply a weaker magnetar. The magnetar interior magnetic field corresponding to Models A and B above are $B_\star \approx ( 6 E_{B_\star}/R_\star^3 )^{1/2} = 1.4 \times 10^{15}\,{\rm G}$ and $8.3 \times 10^{15}\,{\rm G}$, respectively. The nebula predicted by these two models is also $\sim3-10$ times more compact than for FRB 20121102A, and the mean energy per injected particle $\chi$ is somewhat higher.
%(though still consistent with the Baryon loading inferred for the SGR 1806-20 giant magnetar flare; \citealt{Granot06})
Taken together, this modeling---though far from an exhaustive investigation---suggests that the central engine powering the nebula in 20181030A-S1 differs from the source in 20121102A. This indicates a preference towards a system formed with less extreme parameters that never achieved high RM and PRS luminosity, rather than an identical source at a later evolutionary stage.
That being said, it is worth noting that one of the original 20121102A models in \cite{margalitmetzger18} (`Model A' in that work) comes very close to matching both the RM and 1.7 GHz PRS luminosity of 20181030A-S1 at a relatively late epoch of $t \simeq 104\,{\rm yr}$. However the radio spectrum predicted by this model at that epoch is in tension with the broadband spectrum of 20181030A, and in particular over-predicts the 144 MHz flux implied by the LoTSS data.

The EVN upper limits on the compact emission at 5 and 8 GHz presented in \citep{Bruni26} further constrain the physical properties of the nebula. In particular, together with the available low-frequency measurements, they suggest that the spectrum peaks between a few hundred MHz and $\sim1$ GHz, somewhat similarly to what is observed for the PRS associated with FRB 20190417A \citep{Bruno26}, but showing a steeper declining at high frequencies. Within the MWN framework, such behaviour is naturally interpreted as the observed GHz emission probing the exponential cutoff of the injected electron distribution, rather than the shallower synchrotron spectrum produced by cooled electrons injected at earlier epochs. This, in turn, implies additional constraints on the physical properties of the nebula, including an upper limit on the nebular magnetic field,

\begin{equation}
B \lesssim 6\,\mathrm{mG}\,\chi_{0.2}^{-2},
\end{equation}

where $\chi_{0.2}=\chi/(0.2,{\rm GeV})$, and, assuming that the observed RM is dominated by cooled electrons injected early in the source evolution, a corresponding lower limit on the total energy,

\begin{equation}
E \gtrsim 4 \times 10^{47}\,\mathrm{erg}\,
\chi_{0.2}^{3}\,R_{17}^{2},
\end{equation}

where $R_{17}=R/10^{17}\,$cm.

However, a more detailed exploration of these constraints requires a dedicated fit of the model parameter space, which is beyond the scope of the present work.

\subsection{Other possible scenarios for 20181030A-S1}\label{sec: altern}

As already discussed, the localization region of FRB 20181030A hampers the link with 20181030A-S1, hence preventing it to be confirmed as a genuine PRS. In the following, we consider some alternative scenarios to the MWN model discussed in the previous Section. We start by considering alternative PRS models (i.e. scenarios in which 20181030A-S1 resides in NGC 3252 and is physically linked to the FRB source). Additionally, if 20181030A-S1 is not physically associated with FRB 20181030A, two main scenarios can be considered, depending on its distance: (i) the source is located within the host galaxy NGC 3252 but is unrelated to the FRB, or (ii) it is a background radio source projected along the line of sight.

\subsubsection{Radio source in NGC 3252}

\paragraph{HII region.} If 20181030A-S1 belongs to NGC 3252, two main interpretations can be explored. One possibility is that the source is an H\,\textsc{ii} region, in which persistent radio emission would be produced by free-free transitions in a hydrogen plasma surrounding young and massive O and B-type stars \citep[e.g.][]{Terzian65,Vacca96}. These regions are usually classified based on their linear size extent from hyper compact H\,\textsc{ii} regions (UC H\,\textsc{ii}) having size $< 0.05$ pc up to supergiant H\,\textsc{ii} regions with size $> 100$ pc \citep[e.g.][]{Kurtz02}.

Our EVN size constraint of $R < 0.5$ pc, implies that only scenarios involving a compact (hence young) HII region are viable. The spectral luminosity function at $1.4$ GHz of Galactic compact HII regions is reported in \cite{Mascoop21}, showing a completeness spectral luminosity of $\approx 10^{22}$ erg s$^{-1}$ Hz$^{-1}$ and a knee luminosity of $\approx 10^{23}$ erg s$^{-1}$ Hz$^{-1}$, i.e. more than 3 orders of magnitude less than what observed for 20181030A-S1. 

Following \cite{Reines2020}, we compare 20181030A-S1 to the case of W49A, i.e. one of the youngest and brightest HII regions known to reside in the Milky Way. From \cite{Mezger67}, we estimate its spectral luminosity as $\approx 7 \times 10^{24}$ erg s$^{-1}$ Hz$^{-1}$ at $1.7$ GHz, i.e. still 14 times less than 20181030A-S1. While approximately $80$ O-type stars are needed to sustain the radio luminosity of W49A, 20181030A-S1 would require the totality of $\approx 600$ O-type stars\footnote{As in \cite{Reines2020}, we estimate the number of O-type stars needed to sustain the observed radio luminosity as $N_\star = Q_{\rm Lyc}/Q_{\rm Lyc}^{\rm ref}$, where $Q_{\rm Lyc}$ is the rate of Lymann continuum photons, which depends linearly on $L_\nu$ \citep[see][]{Condon92,Reines2020}, and $Q_{\rm Lyc}^{\rm ref} = 10^{49}$ s$^{-1}$ is the reference Lymann rate for a O7.5 V-type star \citep{Vacca96}, which is typically associated with HII regions.}. As already noted in \cite{Ibik24}, the local SFR at the position of 20181030A-S1 is $\approx 0.02\ M_\odot$ yr$^{-1}$, i.e. approximately 2 orders of magnitude less than typical HII regions \citep{Crowther13}. These considerations make a radio source similar to the population of Galactic HII regions unlikely for the origin of 20181030A-S1.

\paragraph{Supernova remnant.} Another possibility is that 20181030A-S1 is a supernova remnant (SNR) in the outskirts of NGC 3252. Following \citet{Reines2020}, we compare the luminosity of 20181030A-S1 with that of Cassiopeia A (Cas A), one of the brightest known SNRs in the Milky Way. Adopting $L_{9} = 7 \times 10^{24}$ erg s$^{-1}$ Hz$^{-1}$ for Cas A at 9 GHz \citep{Reines2020}, we estimate a spectral luminosity at 1.7 GHz of $L_{1.7} \approx 2.5 \times 10^{25}$ erg s$^{-1}$ Hz$^{-1}$, implying that 20181030A-S1 is approximately four times more luminous than Cas A at the same frequency.

We further assess this scenario using the empirical relation between the maximum radio luminosity of SNRs/SNe in a galaxy and its SFR \citep{ChomiukWicots09}:
\begin{equation}
L_{1.4}^{\rm max} = (95^{+31}_{-23})\ {\rm SFR}^{0.98 \pm 0.12},
\end{equation}
with $L_{1.4}^{\rm max}$ in units of $10^{24}$ erg s$^{-1}$ Hz$^{-1}$. Using the total SFR of NGC 3252, $\log_{10}(\mathrm{SFR}) = -0.45 \pm 0.1$ \citep{Bhardwaj21}, we obtain an expected maximum luminosity at $1.4$ GHz of $L_{1.4}^{\rm max} \approx 3.4 \times 10^{25}$ erg s$^{-1}$ Hz$^{-1}$. Even accounting for the intrinsic scatter in the relation, this value remains well below the luminosity observed for 20181030A-S1 by a factor of $\approx 3$. 

The upper limit on the radius ($R \lesssim 0.5$ pc) implies also that, it would have to be either exceptionally young or expanding into an unusually dense ambient medium. Assuming that the remnant has already entered the Sedov--Taylor phase, the size constraint implies \citep[e.g.,][]{Sedov59, Draine11}

\begin{equation}
\left(\frac{t}{100\,\mathrm{yr}}\right)^{2/5}
\left(\frac{n}{10^{3}\,\mathrm{cm}^{-3}}\right)^{-1/5}
\lesssim 1,
\end{equation}

where $t$ is the age of the remnant and $n$ is the ambient ISM density. Even adopting an extreme density of $n \sim 10^{3}$ cm$^{-3}$, the remnant would need to be younger than $\sim100$ yr. Such a young, compact, and radio-luminous SNR would be expected to be exceptionally rare, making this interpretation unlikely. Finally, the observed emission is unlikely to arise from a population of unresolved SNRs. Given the upper limit on the source size ($R\lesssim0.5$ pc), multiple remnants would have to be confined within a region of order 1 pc to remain unresolved by our EVN observations, which appears highly implausible.

\subsubsection{Background compact radio source}
Finally, an alternative scenario is that 20181030A-S1 is a background compact radio source unrelated to both FRB 20181030A and NGC 3252. In this case, its apparent association with the FRB localization region and the host galaxy would be due to a chance alignment. \cite{Ibik24} estimate $P_{\rm cc} \approx 0.05$ as the chance-coincidence probability of associating a radio source with flux density greater than 20181030A-S1 to NGC 3252. While this value is relatively small, it is not sufficiently low to rule out a chance alignment.

Assuming a typical radio luminosity of $L_\nu \sim 10^{29}$--$10^{31}$ erg s$^{-1}$ Hz$^{-1}$ for core-dominated AGN, the observed flux density of $\sim 280\ \mu$Jy at 1.7 GHz would place the source at a cosmological distance (i.e., $z \sim 0.5$--$1$, depending on the assumed luminosity and neglecting $k$-corrections). However, the radio spectrum also disfavors this interpretation. As discussed in Sect.~5, the non-detections at 5 and 8 GHz imply an upper limit on the spectral index of $\alpha \lesssim -1.1$ (where $F_\nu \propto \nu^\alpha$). Such a steep spectrum is difficult to reconcile with the flat radio spectra typically observed in the compact cores of radio-loud AGN. For example, BL Lac objects generally exhibit spectral indices between 1.7 and 5 GHz close to $\alpha \sim 0$, with a typical scatter of only $\sim 0.1$ \citep[e.g.,][]{Giroletti06}.

We searched for an optical counterpart co-located with 20181030A-S1 in a Pan-STARRS g filter image but found none. The limiting magnitude is $m_g \simeq 23.3$ \citep{Chambers16}. However, the bright diffuse emission from the outskirts of NGC~3252 significantly limits the sensitivity at the source position, preventing stringent constraints on the presence of a faint optical counterpart. Further constraints on the nature of the source will require improved spectral measurements, as well as deeper and higher-resolution optical imaging \citep[see, e.g., the Euclid Wide Survey;][]{Euclid22} to search for a potential background host galaxy.

\section{Summary and conclusions}\label{sec: conclusions}

In this work, we presented new EVN observations at $1.7$ GHz of 20181030A-S1, the candidate persistent radio source (PRS) associated with the repeating FRB 20181030A \citep{chime18,Ibik24}. This system is particularly compelling given the proximity of its proposed host galaxy, NGC 3252 \citep{Bhardwaj21}, at a luminosity distance of $20 \pm 5$ Mpc \citep{Tully16}. Our observations reveal a compact radio source, with an upper limit on its size of $<0.5$ pc ($1\sigma$), assuming it lies at the distance of NGC 3252. If confirmed as a genuine PRS, this would make 20181030A-S1 the nearest such source known, compared to the previously closest example associated with FRB 20190417A at $z = 0.12817(2)$ \citep[$\approx 600$ Mpc;][]{Ibik24,Moroianu25}.

We measure a flux density of $280 \pm 30\ \mu$Jy at $1.7$ GHz, corresponding to a spectral luminosity of $9(1) \times 10^{25}$ erg s$^{-1}$ Hz$^{-1}$ at 20 Mpc. This is $2$--$3$ orders of magnitude lower than the luminosities of known PRSs, making 20181030A-S1, if associated with the FRB, the faintest PRS identified so far. Even accounting for a $\sim 15\%$ uncertainty in the absolute flux scale, our measurement remains below the $400 \pm 10\ \mu$Jy reported at arcsecond resolution \citep{Ibik24}, suggesting that the latter likely includes an extended emission component, while our EVN observations isolate the compact core.

If 20181030A-S1 is physically associated with FRB 20181030A, its observed properties can be interpreted within the framework of a MWN. By modeling the evolution of the nebular emission and rotation measure following \citet{margalitmetzger18}, we find that physically plausible nebula models can simultaneously reproduce the observed PRS spectral luminosity, the measured RM of the FRB, and the compact size constraint derived from our EVN observations. Although our modeling is not intended as a formal parameter inference, it suggests that 20181030A-S1 could be powered by a less extreme central engine than the PRS associated with FRB 20121102A, producing a more compact and less luminous nebula, somewhat different with respect the prototypical FRB-PRS system associated with R1. The recent EVN upper limits at 5 and 8 GHz \citep{Bruni26} further suggest a spectral peak between a few hundred MHz and $\sim1$ GHz, providing additional constraints on the particle distribution and magnetic field of the nebula. 

It is interesting to note that the apparently less extreme properties inferred for the putative central engine of 20181030A-S1 may also be qualitatively consistent with the different environment of the host galaxy with respect to confirmed PRSs. If the strength of the magnetic field generated in newborn magnetars is linked to the rotation rate and/or mass of their progenitor stars, as suggested in some formation scenarios \citep[see, e.g.,][]{Metzger17,Margalit18}, then the higher metallicity of NGC~3252 compared to the low-metallicity hosts of confirmed PRSs\footnote{The gas-phase metallicity of NGC~3252 is estimated to be 
$12+\log(\mathrm{O/H})=8.44\pm0.06$ \citep{Bhardwaj21}, compared to 
$12+\log(\mathrm{O/H})=8.0\pm0.1$ for the host galaxy of FRB~20121102A 
\citep{Bassa17} and $12+\log(\mathrm{O/H})=7.95\pm0.01$ for the host 
galaxy of FRB~20190417A \citep{Moroianu25}.} could lead to less rapidly rotating progenitors and, consequently, to less strongly magnetized neutron stars. Such a central engine would be expected to power a weaker and more slowly evolving MWN, qualitatively consistent with the low luminosity inferred for 20181030A-S1. Although this interpretation remains speculative, since FRB host galaxy do not exhibit a strong metallicity preference  \citep{Yamasaki25}, it suggests that the properties of FRB-associated PRSs may depend not only on age but also on the environment in which the magnetar forms \citep[see also][]{Moroianu25}, a hypothesis that can be tested through future population studies of well-characterized FRB--PRS systems, as their sample size increases.

We also explored alternative scenarios for the origin of the radio source, given the large uncertainties in the position of FRB 20181030A. We show that, under the hypothesis that 20181030A-S1 is located within NGC 3252 but unrelated to the FRB, its observed properties are difficult to reconcile with known classes of compact radio emitters powered by star formation. In particular, an H\,\textsc{ii} region interpretation would require an extreme concentration of massive stars ($\sim 600$ O-type stars) within a sub-parsec region, far exceeding expectations based on the local star formation rate. Similarly, an origin as an individual supernova remnant is strongly disfavoured by both the flat-to-inverted radio spectrum and the observed luminosity, which exceeds the expected maximum SNR luminosity for a galaxy like NGC 3252.

Alternatively, 20181030A-S1 could be a background compact radio source, such as a core-dominated AGN, whose flat spectrum would be naturally explained by partially self-absorbed synchrotron emission. In this scenario, the observed flux density would imply a cosmological distance ($z \sim 0.5$--$1$), and the apparent association with NGC 3252 would arise from a chance alignment. Although the estimated chance-coincidence probability ($P_{\rm cc} \approx 0.05$; \citealt{Ibik24}) is relatively small, it is not negligible, and this possibility cannot be excluded.

Overall, if 20181030A-S1 is indeed a genuine PRS associated with FRB 20181030A, it demonstrates that persistent radio emission linked to FRBs can exist at luminosities well below previously established limits, while remaining compact and physically connected to the central engine. This would imply that PRSs may be far more common than currently inferred, with most lying below the detection thresholds of existing surveys.

Future observations will be crucial to clarify the nature of 20181030A-S1. In particular, improved FRB localizations, broadband radio spectral measurements, and deeper high-resolution optical imaging will be key to distinguishing between a nearby, faint PRS and a background compact radio source.

\begin{acknowledgements}
The European VLBI Network is a joint facility of independent European, African, Asian, and North American radio astronomy institutes. Scientific results from data presented in this publication are derived from the EP134 EVN project code. The research activities described in this paper were carried out with contribution of the NextGenerationEU funds within the National Recovery and Resilience Plan (PNRR), Mission 4 - Education and Research, Component 2 - From Research to Business (M4C2), Investment Line 3.1 - Strengthening and creation of Research Infrastructures, Project IR0000026 – Next Generation Croce del Nord. LOFAR is the Low Frequency Array designed and constructed by ASTRON. It has observing, data processing, and data storage facilities in several countries, which are owned by various parties (each with their own funding sources), and which are collectively operated by the LOFAR ERIC under a joint scientific policy. The LOFAR resources have benefited from the following recent major funding sources: CNRS-INSU, Observatoire de Paris and Université d'Orléans, France; BMFTR, MKW-NRW, MPG, Germany; Science Foundation Ireland (SFI), Department of Business, Enterprise and Innovation (DBEI), Ireland; NWO, The Netherlands; The Science and Technology Facilities Council, UK; Ministry of Science and Higher Education, Poland; The Istituto Nazionale di Astrofisica (INAF), Italy. This research made use of the Dutch national e-infrastructure with support of the SURF Cooperative (e-infra 180169) and the LOFAR e-infra group. The Jülich LOFAR Long Term Archive and the German LOFAR network are both coordinated and operated by the Jülich Supercomputing Centre (JSC), and computing resources on the supercomputer JUWELS at JSC were provided by the Gauss Centre for Supercomputing e.V. (grant CHTB00) through the John von Neumann Institute for Computing (NIC). This research made use of the University of Hertfordshire high-performance computing facility and the LOFAR-UK computing facility located at the University of Hertfordshire and supported by STFC [ST/P000096/1], and of the Italian LOFAR-IT computing infrastructure supported and operated by INAF, including the resources within the PLEIADI special "LOFAR" project by USC-C of INAF, and by the Physics Department of Turin university (under an agreement with Consorzio Interuniversitario per la Fisica Spaziale) at the C3S Supercomputing Centre, Italy. This research is part of the project LOFAR Data Valorization (LDV) [project numbers 2020.031, 2022.033, and 2024.047] of the research programme Computing Time on National Computer Facilities using SPIDER that is (co-)funded by the Dutch Research Council (NWO), hosted by SURF through the call for proposals of Computing Time on National Computer Facilities. CS acknowledges financial support by the Italian Ministry of University and Research (grant FIS2023$-$01611, CUP C53C25000300001) and by the INAF (through \textsl{Ricerca Fondamentale 2024}, Ob. Fu. 1.05.24.07.04).

\end{acknowledgements}

\bibliographystyle{aa}
\bibliography{biblio}

\begin{appendix}
\nolinenumbers
\section{LoTSS and VLASS image}
\begin{figure}
    \centering
    \includegraphics[width=1.0\columnwidth]{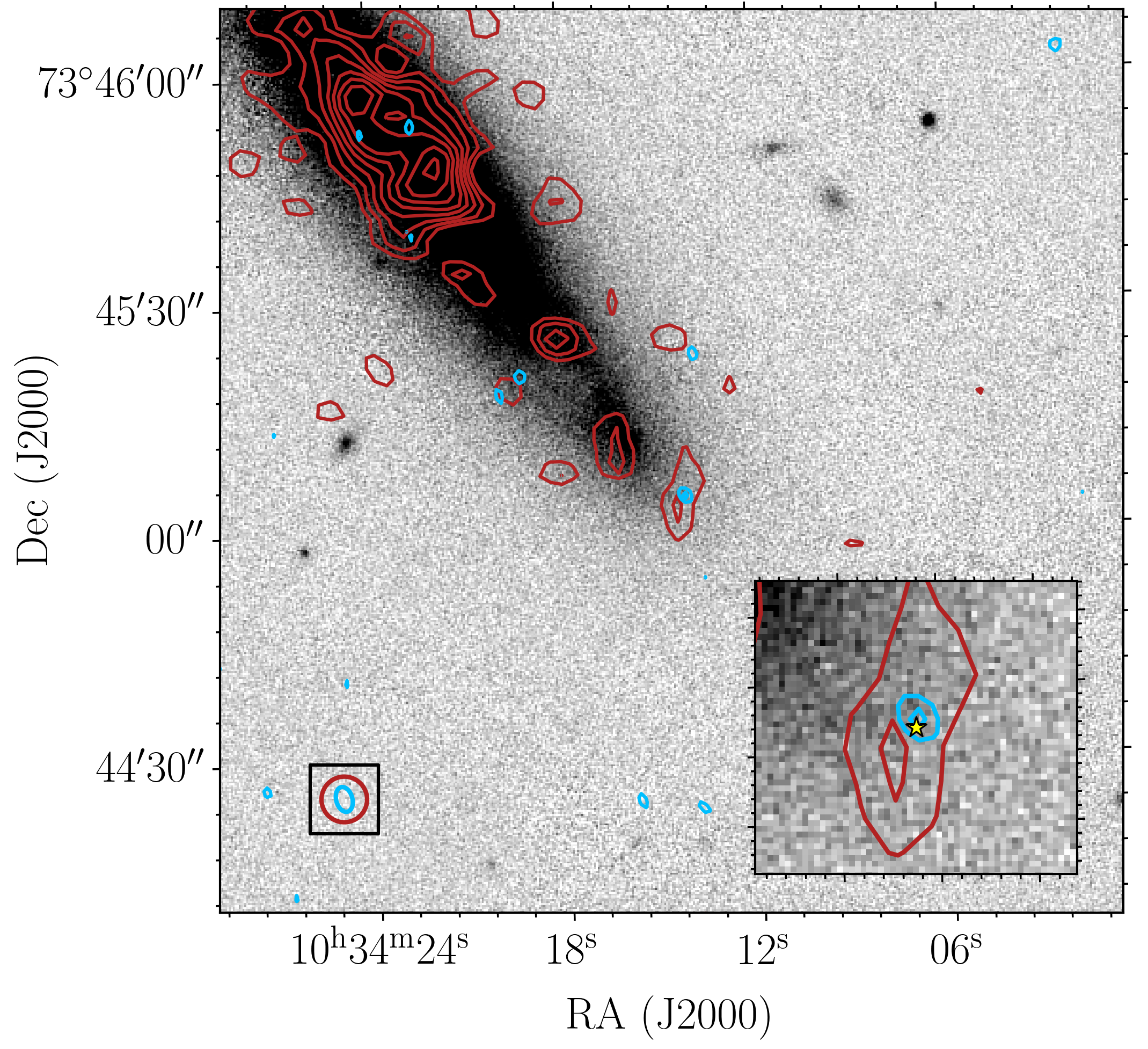}
    \caption{Pan-STARRS $z$-band optical image of the field around 20181030A-S1. The radio emission detected by LoTSS-DR3 and VLASS is overlaid with red and blue contours, respectively. The contour levels range from three to ten times the rms noise level of each radio image, i.e. $\sigma = 72$ $\mu$Jy beam$^{-1}$ and $\sigma = 118$ $\mu$Jy beam$^{-1}$ for LoTSS and VLASS, respectively. The LoTSS and VLASS synthesized beams are shown as unfilled ellipses in the bottom-left corner of the image. The inset provides a zoom-in of the region around 20181030A-S1, whose position is marked by a yellow star.}
    \label{fig:LoTSS_VLASS}
\end{figure}

In Fig. \ref{fig:LoTSS_VLASS} we show the field around the PRS candidate 20181030A-S1, with overlaied contours from LoTSS-DR3 and VLASS. The position of the PRS candidate is indicated with a yellow star in the Figure inset.

\section{Likelihood function and parameterisation}
\label{app:A}

We model the relation between the radio luminosity and the Faraday rotation measure by assuming a log-normal scatter in $L_\nu$ at fixed RM. Working in logarithmic\footnote{In this paper, all logarithms are base 10; we write $\log$ for brevity.} space, we define the observables as $y_i \equiv \log L_{\nu,i}$ and the model prediction as $\hat{y}_i(\boldsymbol{\theta})$, where $\boldsymbol{\theta}$ denotes the set of model parameters. In this work, the model is parameterized by a normalisation parameter $A \equiv \zeta_e \gamma_{\rm th}^2 (R/0.01\ {\rm pc})^2$, so that Eq. \ref{eq: RM_Lnu} can be written as
\begin{equation}
\hat{y}_i(\boldsymbol{\theta}) = \log A + \log \mathrm{RM}_i\ .
\end{equation}

We assume that the logarithmic luminosities are drawn from a Gaussian distribution centered on the model prediction, with a variance given by the sum of the observational uncertainty and an intrinsic scatter term:
\begin{equation}
\log L_{\nu,i} \sim \mathcal{N}\!\left(\hat{y}_i(\boldsymbol{\theta}), \sigma_{{\rm tot},i}^2 \right),
\end{equation}
with
\begin{equation}
\sigma_{{\rm tot},i}^2 = \sigma_{y,i}^2 + \sigma_{\rm int}^2,
\end{equation}
where $\sigma_{y,i}$ is the measurement uncertainty on $\log L_{\nu,i}$ and $\sigma_{\rm int}$ is a free parameter accounting for intrinsic dispersion in the population.

Under these assumptions, the log-likelihood for a dataset of $N$ sources reads
\begin{equation}
\ln \mathcal{L}(\boldsymbol{\theta}) =
-\frac{1}{2} \sum_{i=1}^{N}
\left[
\frac{\left(\log L_{\nu,i} - \hat{y}_i(\boldsymbol{\theta})\right)^2}{\sigma_{{\rm tot},i}^2}
+ \ln\!\left(2\pi \sigma_{{\rm tot},i}^2\right)
\right].
\end{equation}

The intrinsic scatter $\sigma_{\rm int}$ enters only through the likelihood function and is treated as a nuisance parameter. After verifying the convergence of the MCMC sampling, we marginalise over $\sigma_{\rm int}$ when deriving constraints on the remaining model parameters.

Finally, instead of sampling directly the normalisation parameter $A$, we perform the inference in terms of $\log A$. This choice is motivated by both numerical efficiency and prior considerations. In particular, the physically motivated prior range for $A$ spans several orders of magnitude, $A \in [10^{-5}, 10^{2}]$. Sampling in linear space would lead to a poor exploration of the low-$A$ region and inefficient convergence of the chains. By sampling $\log A$ with a uniform prior, we ensure a more homogeneous exploration of the parameter space and a better sampling of small values of $A$, improving the stability and convergence properties of the inference.

\end{appendix}

\FloatBarrier

\end{document}